\documentclass[12pt]{iopart}
\usepackage{graphicx}
\usepackage{epstopdf}
\usepackage{amssymb}

\usepackage{color}
\usepackage[round]{natbib}
\newcommand{\newblock}{}
\graphicspath{{figures/}}
\usepackage{tabularx}
\usepackage{multirow}
\usepackage{booktabs}

\usepackage{iopams}
\usepackage{changepage}
  \expandafter\let\csname equation*\endcsname\relax
  \expandafter\let\csname endequation*\endcsname\relax
\usepackage[cmex10]{amsmath}
\newcommand{\beq}{\begin{equation}}
\def\eeq{\end{equation}}

\begin{document}
\bibliographystyle{natbib} %>>>>

\title[Improved Image-Domain Material Decomposition]{Using Edge-Preserving Algorithm with Non-local Mean for  Significantly Improved Image-Domain Material Decomposition in Dual Energy CT}

\author{Wei Zhao$^{1}$,~Tianye~Niu$^{2}$,~Lei Xing$^{3}$,Yaoqin Xie$^{4}$,~Guanglei Xiong$^{5}$,~Kimberly Elmore$^{5}$,~Jun Zhu$^{1}$, Luyao Wang$^{1}$,~James K.~Min$^{5}$}%
\address{$^1$ Department of Biomedical Engineering, Huazhong University of Science and Technology, Hubei, 430074 China}
\address{$^2$ Sir Run Run Shaw Hospital, Zhejiang University School of Medicine; Institute of Translational Medicine, Zhejiang University, Hangzhou, Zhejiang, 310016 China}
\address{$^{3}$ Stanford University, Department of Radiation Oncology, Stanford, CA 94305 USA}
\address{$^{4}$ Shenzhen Institutes of Advanced Technology, Chinese Academy of Sciences, Shenzhen 518055 China}
\address{$^{5}$ Dalio Institute of Cardiovascular Imaging New York-Presbyterian Hospital and Weill Cornell Medical College, New York, NY 10021}

\ead{jkm2001@med.cornell.edu}

\begin{abstract}

\textcolor{black}{Increased} noise is a general concern for dual-energy material decomposition. Here, we develop an image-domain material decomposition algorithm for dual-energy CT (DECT) by incorporating an edge-preserving filter into the Local HighlY constrained backPRojection Reconstruction (HYPR-LR) framework. With effective use of the non-local mean, the proposed algorithm, which is referred to as HYPR-NLM, reduces the noise in dual energy decomposition while preserving the accuracy of quantitative measurement and spatial resolution of the material-specific dual energy images. We demonstrate the noise reduction and resolution preservation of the algorithm with iodine concentrate numerical phantom by comparing the HYPR-NLM algorithm to the direct matrix inversion, HYPR-LR and iterative image-domain material decomposition (Iter-DECT). We also show the superior performance of the HYPR-NLM over the existing methods by using two sets of cardiac perfusing imaging data. The DECT material decomposition comparison study shows that all four algorithms yield acceptable quantitative measurements of iodine concentrate. Direct matrix inversion yields the highest noise level, followed by HYPR-LR and Iter-DECT. HYPR-NLM in iterative formulation significantly reduces image noise and the image noise is comparable to or even lower than that generated using Iter-DECT. For the HYPR-NLM method, there are marginal edge effects in the difference image, suggesting the high-frequency details are well preserved. In addition, when the search window size increases from $11\times11$ to $19\times19$ , there are no significant change and marginal edge effects in the HYPR-NLM difference images. The reference drawn from the comparison study includes: (1) HYPR-NLM significantly reduces the DECT material decomposition noise while preserving quantitative measurements and high-frequency edge information, and (2) HYPR-NLM is robust with respect to parameter selection. %Additionally, comparison studies were retrospectively performed for two sets of cardiac perfusing imaging data.

\end{abstract}

%\pacs{ 87.57.Q- Computed tomography, 87.59.-e X-ray imaging, 87.57.cf Spatial resolution, 87.57.C- Image quality}

\maketitle

%%%%%%%%%%%%%%%%%%%%%%%%%%%%%%%%%%%%%%%%%%%%%%%%%%%%%%%%
\section{Introduction}

In dual-energy CT imaging (DECT), the object is scanned using two energy spectra with different kVp settings, or different prefiltration, or both. Compared to standard CT imaging where only one x-ray spectrum was used to yield an effective linear attenuation coefficient of the object, DECT can take advantage of the energy dependence of the linear attenuation coefficients, yielding energy and material-selective images~\citep{alvarez1976,kalender1986}. This enables DECT to be used in various clinical applications, including improving tissue or contrast agent segmentation and quantification~\citep{johnson2007,liu2009,fischer2011,chandarana2011,li2013}, removing beam hardening artifacts~\citep{wu2009,yu2011,yu2012,scheske2013}, and further providing improved clinical significance for modern CT scanners~\citep{graser2009,hartman2012,marin2014}.

The material-selective image is obtained by dual-energy basis material decomposition, which depends on the mass attenuation coefficients of the basis materials as well as dual-energy projection data acquisition. Dual-energy raw data can be acquired in several different ways, such as sequential scans at different kVps, dual x-ray sources at $90^{\circ}$ on the same gantry~\citep{johnson2007}, fast kVp switching within a single scan~\citep{kalender1986,silva2011,matsumoto2011} and consistent ray approach using either layered detectors~\citep{carmi2005,hao2013} or photon counting detectors~\citep{wang2009,wang2010,shikhaliev2012}. Depending on the dual-energy data acquisition method, basis material decomposition can be performed either in projection-domain~\citep{sidky2004,stenner2007,noh2009,brendel2009,maass2011,xing2013} or image-domain~\citep{maass2009,niu2014,clark2014,li2015,faby2015,petrongolo2015}, or joint-domain~\citep{sukovic2000,zhang2014,long2014}. In this study, we will focus on image-domain material decomposition.

In the theory of image-domain material decomposition, the linear attenuation coefficients derived from reconstructed images at low- and high-energy scans can be expressed as a linear combination of the pixel values in the images of the two basis materials~\citep{kalender1986,szczykutowicz2010,niu2014}:
\beq\label{equ:decomEquation}
{\mu_{H} \choose \mu_{L}} =
\begin{pmatrix}
  \mu_{1H} & \mu_{2H}  \\
  \mu_{1L} & \mu_{2L}
 \end{pmatrix}
 {x_{1} \choose x_{2}},
\eeq
where $\mu_{H,L}$ \textcolor{black}{($cm^{-1}$)} is the measured attenuation coefficient of a specific pixel using high ($H$) and low ($L$) energy spectrum, and \textcolor{black}{the unitless} $x_{1,2}$ \textcolor{black}{denote} the projection component of the attenuation coefficient along two basis materials 1 and 2, respectively. \textcolor{black}{After taking mass conservation into account, $x_{1,2}$ can be regarded as volume fractions.} $\mu_{ij}$ \textcolor{black}{($cm^{-1}$)} is the linear attenuation coefficient of material $i$ ($i=1$ or 2) under the energy spectrum $j$ ($j=H$ or $L$). The linear attenuation coefficients $\mu_{ij}$ of the basis materials can be either obtained from the ROIs in the high and low energy images that correspond to the basis materials or from the calibration scan using the basis material phantom.% that are made with the basis materials. The subscripts 1 and 2 stand for the two material bases, and

\textcolor{black}{In cases DECT raw data acquired using two source-detector pairs that are using different spectra, and fast kVp switching during a single source scan, the high- and low-energy CT scans are geometrically inconsistent (i.e., the paths of high- and low-energy CT measurements are not based on the same path) and some dedicated algorithms are designed to deal with this problem~\citep{knaup2007,maass2011}}. Compared to projection-domain decomposition methods which may suffer from inconsistent rays issue, image-domain dual-energy decomposition is more convenient in clinical applications as it is performed on reconstructed CT images acquired on commercial CT scanners. In this case, material-specific images can be simply generated using direct matrix inversion, however, direct material decomposition techniques such as matrix inversion \textcolor{black}{can} yield \textcolor{black}{amplified} image noise \textcolor{black}{since the low- and high-energy signals are subtracted while the noise is the summation of the two.} Note that \textcolor{black}{the amplified} noise is a common issue for both projection and image-domain material decomposition. Many \textcolor{black}{methods} have been proposed to address the problem~\citep{maass2009,shen2013,clark2014,dong2014,petrongolo2015}.

This study aims to significantly improve DECT imaging by establishing a new theoretical framework of image-domain material decomposition with incorporation of  edge-preserving techniques.  We demonstrate the advantages of the proposed approach by digital phantom studies and by  quantification of iodine concentration of a series of myocardial perfusion imaging studies.

\section{Methods and materials}
\vspace{-0.5mm}
\subsection{Material decomposition via direct matrix inversion}

Suppose the total number of pixels of one CT image is $N$, for low- and high-energy CT images and material-specific images, equation~(\ref{equ:decomEquation}) can be rewritten in matrix form,
\beq\label{equ:MatrixForm}
\vec{\mu}=A\vec{x}.
\eeq
Here $A$ is a $2N\times2N$ material decomposition matrix and it can be derived from equation~(\ref{equ:decomEquation}) as,
\beq\label{equ:matrix}
A=
\begin{pmatrix}
  \mu_{1H}I & \mu_{2H}I  \\
  \mu_{1L}I & \mu_{2L}I
 \end{pmatrix},
\eeq
with $I$ the $N\times N$ identity matrix. $\vec{\mu}$ and $\vec{x}$ are column \textcolor{black}{vectors} with dimension of $2N$ as \textcolor{black}{they consist} of high- and low-energy CT image, and decomposed material-specific images with column vector form, respectively. Namely,
\beq\label{equ:matrix}
\vec{\mu}=
\begin{pmatrix}
  \vec{\mu}_H\\
  \vec{\mu}_L
 \end{pmatrix},~~
  \vec{x}=
\begin{pmatrix}
  \vec{x}_1\\
  \vec{x}_2
 \end{pmatrix},
\eeq
with $\vec{\mu}_H$ and $\vec{\mu}_L$ the measured high- and low-energy CT images, respectively, and $\vec{x}_1$ and $\vec{x}_2$ the two material-specific images. Note that both $\vec{\mu}_{H,L}$ and $\vec{x}_{1,2}$ are represented as vectors.
To obtain material-specific images from high- and low-energy CT images, one can directly use matrix inversion to solve equation~(\ref{equ:decomEquation}), i.e.
\beq\label{equ:dirInverse}
\vec{x}=A^{-1}\vec{\mu},
\eeq
where
\beq\label{equ:matrixInA}
A^{-1}=C
\begin{pmatrix}
  -\mu_{2L}I & \mu_{2H}I\\
  \mu_{1L}I & -\mu_{1H}I
 \end{pmatrix}.
\eeq
 with \textcolor{black}{matrix determinant} $C=(\mu_{2H}\mu_{1L}-\mu_{1H}\mu_{2L})^{-1}$. However, material decomposition in this manner \textcolor{black}{can} yields significantly \textcolor{black}{increased} noise and severely degraded noise-to-signals (NSRs) for the material-specific images. A method taking some measures to reduce the noise and to recover NSRs is highly desirable.

\subsection{HYPR-NLM}

To reduce the \textcolor{black}{amplified} image noise in the decomposed material-specific images while keeping the images as accurate as possible (both quantitative measurement and spatial resolution), we developed an image-domain material images denoising algorithm. The algorithm \textcolor{black}{was} based on the HYPR-LR (Local HighlY constrained backPRojection Reconstuction) framework~\citep{mistretta2006,johnson2008,leng2011}, which was first proposed for reconstruction of time-resolved magnetic resonance imaging (MRI) using highly undersampled projection. The HYPR-LR framework has been successfully applied to a broad range of medical imaging applications, such as dynamic MRI~\citep{johnson2008}, CT angiography~\citep{supanich2009}, dynamic positron emission tomography~\citep{christian2010}, spectral CT~\citep{leng2011} and myocardial perfusion imaging~\citep{speidel2013}. For noise reduction in spectral CT, the HYPR-LR algorithm \textcolor{black}{treated} CT images at different energies as four-dimensional CT images, i.e., the energy dimension \textcolor{black}{was} regarded as time dimension. It \textcolor{black}{started} with producing a composite image $\vec{\mu}_c$ by averaging the energy images \textcolor{black}{with equal weighting}, i.e. yielding an image with lower image noise level by using the images from different energy bins of photon-counting detector CT, or from high- and low-energy images of dual-energy CT. A weight image \textcolor{black}{was} then generated by calculating the ratio of two filtered images, which \textcolor{black}{were} obtained by filtering both the energy images $\vec{\mu}_{e}$ and the composite image $\vec{\mu}_c$. The final image $\vec{\mu}_{he}$ \textcolor{black}{was} obtained by multiplying the weight image and the composite image. The mathematical form of HYPR-LR algorithm is as follows,
\beq\label{equ:HYPR-LR}
\vec{\mu}_{he} = \frac{\vec{\mu}_{e}\otimes K}{\vec{\mu}_c \otimes K} \cdot \vec{\mu}_c
\eeq
with $\vec{\mu}_{e}$ the image with different energy, and $K$ the low-pass filter kernel (usually an uniform kernel). The symbol $\otimes$ stands for convolution operation. For dual-energy material decomposition, HYPR-LR has been applied to the high- and low-energy CT images to yield noised reduced images, as well as superior material-specific basis images~\citep{leng2011}.%To our knowledge, see Appendix

In this study, we \textcolor{black}{demonstrated} the HYPR-LR algorithm can be applied directly to the dual energy material-specific images with composite image $\vec{\mu}_c$.
For dual-energy high- and low- energy images $\vec{\mu}_{H,L}$, the processed \textcolor{black}{individual energy} images $\vec{\mu}_{he,H}$ and $\vec{\mu}_{he,L}$ using the HYPR-LR algorithm are calculated as follows,
\beq\label{equ:HYPR-LRh}
\vec{\mu}_{he,H} = \frac{\vec{\mu}_{H}\otimes K}{\vec{\mu}_c \otimes K} \cdot \vec{\mu}_c,
\eeq
\beq\label{equ:HYPR-LRl}
\vec{\mu}_{he,L} = \frac{\vec{\mu}_{L}\otimes K}{\vec{\mu}_c \otimes K} \cdot \vec{\mu}_c.
\eeq
Based on equation~(\ref{equ:dirInverse}), the basis material image $\vec{x}_{h1}$ calculated using $\vec{\mu}_{he,H}$ and $\vec{\mu}_{he,L}$ is expressed as,
\begin{eqnarray}\label{equ:HYPR-LRb1}
\begin{split}
\vec{x}_{h1}=& -C\left(\mu_{2L}\vec{\mu}_{he,H}-\mu_{2H}\vec{\mu}_{he,L}\right) \\
=&-C \left( \mu_{2L} \frac{\vec{\mu}_{H}\otimes K}{\vec{\mu}_c \otimes K} \cdot \vec{\mu}_c - \mu_{2H}\frac{\vec{\mu}_{L}\otimes K}{\vec{\mu}_c \otimes K} \cdot \vec{\mu}_c \right)\\
=&-C\left( \frac{(\mu_{2L}\vec{\mu}_{H}-\mu_{2H}\vec{\mu}_{L}) \otimes K}{\vec{\mu}_c \otimes K} \cdot \vec{\mu}_c\right)\\
=&\frac{\vec{x}_1 \otimes K}{\vec{\mu}_c \otimes K} \cdot \vec{\mu}_c.
\end{split}
\end{eqnarray}
The above derivation has used the algebraic properties of convolution, i.e., the distributivity and the associativity of convolution with scalar multiplication. Following the fashion in equation~(\ref{equ:HYPR-LRb1}), we have the same expression for the other basis material image $\vec{x}_{h2}$.
%\beq\label{equ:HYPR-LRb2}
%\vec{x}_{h2} = \frac{\vec{x}_2 \otimes K}{\vec{\mu}_c \otimes K} \cdot \vec{\mu}_c.
%\eeq
Equation~(\ref{equ:HYPR-LRb1}) demonstrates the HYPR-LR algorithm can be applied directly to the material-specific image. %The detail of derivation is in the Appendix.
Thus we have,
\beq\label{equ:HYPR-LRb}
\vec{x}_{hb} = \frac{\vec{x}_{b}\otimes K}{\vec{\mu}_c \otimes K} \cdot \vec{\mu}_c.
\eeq
Here $\vec{x}_{b}$ is the material-specific basis image, i.e. $\vec{x}_{1}$ or $\vec{x}_{2}$, and $\vec{x}_{hb}$ is the processed basis image.
 In order to further reduce decomposed images noise without loss of high-frequency edge information, an edge-preserving non-local mean (NLM)~\citep{buades2005} \textcolor{black}{was} introduced into the HYPR framework. Edge-preserving techniques \textcolor{black}{were} usually applied to yield noise reduced, spatial resolution well preserved images~\citep{chen2008a,wangjin2009,zhu2009,chen2012,chun2014}. In this study, NLM \textcolor{black}{was} employed to generate the weight image for the HYPR framework and the resulting algorithm \textcolor{black}{was} referred as HYPR-NLM, hence equation~(\ref{equ:HYPR-LRb}) \textcolor{black}{was} rewritten in pixel-wise fashion as,
\beq\label{equ:HYPR-NLM}
x_{hb}(i) = \frac{\underset{j\in \Omega_i} \sum \omega(i,j)x_{b}(j)}{\underset{j\in \Omega_i} \sum \omega(i,j) \mu_c(j)} \mu_c(i),
\eeq
where the weight $\omega(i,j)$ depicts the similarity between the pixels $i$ and $j$, \textcolor{black}{and it satisfies the constraint conditions $0\leq \omega(i,j) \leq 1$ and $\underset{j} \sum \omega(i,j)=1$ (as demonstrated later)}. The pixel dependent summation domain $\Omega_i$ denotes a search-window centered at the pixel $i$ and it is usually a square neighborhood with fixed size. Thus each pixel of the filtered image is a weighted summation of a square neighborhood. \textcolor{black}{For the right hand side of equation~(\ref{equ:HYPR-NLM}), $\omega(i,j)$ is applied to the material images in the numerator, and to the composite image in the denominator.}.

The similarity between pixels $i$ and $j$ can be measured using a weighted Euclidean distance of two square neighborhoods centered at pixels $i$ and $j$, thus the weight $\omega(i,j)$ is defined as,
\beq\label{equ:weight}
\omega(i,j) = \frac{1}{Z(i)}exp\left(-\frac{\|\vec{\mu}(\Theta_i)-\vec{\mu}(\Theta_j)\|^2_{2,a}}{d^2}\right),
\eeq
where $\Theta_i$ and $\Theta_j$ are two square neighborhoods centered at pixels $i$ and $j$, respectively. The parameter $d$ controls the decay of the exponential function and it acts as a degree of filtering. The parameter $a$ is the standard deviation (SD) of a Gaussian kernel \textcolor{black}{which has the same size as the square neighborhood. During numerical implementation, the Gaussian kernel gives weights to the Euclidean distance of the two square neighborhoods}. $Z(i)$ is the normalization constant calculated as,
\beq\label{equ:weight}
Z(i) = \underset{j} \sum exp\left(-\frac{\|\vec{\mu}(\Theta_i)-\vec{\mu}(\Theta_j)\|^2_{2,a}}{d^2}\right).
\eeq
Based on the above definition, $\omega(i,j)$ satisfies the constraint conditions.
%\textcolor{black}{During numerical implementation, a Gaussian kernel with SD $a$ and size as the same as the square neighborhood is applied to the Euclidean distance of the two square neighborhoods and plays the role of weighting}.

\subsection{Iterative image-domain material decomposition}

\textcolor{black}{Since dual-energy material decomposition can yield noise amplified images, for comparison, we also \textcolor{black}{used} an iterative image domain material decomposition method which significantly \textcolor{black}{reduced} the noise of the material-specific images with superior performance on image spatial resolution and low-contrast detectability~\citep{niu2014}. The method \textcolor{black}{balanced} the data fidelity of the value of the material image and a quadratic penalty using an optimization framework, and the unconstrained optimization problem \textcolor{black}{was} solved by the nonlinear conjugate gradient method (see the Appendix for details). In this work, we \textcolor{black}{referred} this method as Iter-DECT.}

\subsection{Numerical simulation studies}

To evaluate the proposed HYPR-NLM dual-energy material decomposition algorithm, we first \textcolor{black}{used} simulated phantom data. For all of the simulation studies, the high- and low-energy CT images \textcolor{black}{were} first employed to generate the material-specific basis images, which \textcolor{black}{were} then processed using HYPR-NLM method to yield noise reduced images. These images \textcolor{black}{were} compared to the images generated using direct matrix inversion, Iter-DECT and HYPR-LR methods and the results \textcolor{black}{were} quantitatively analyzed.% in image-domain.

In all of the simulations, \textcolor{black}{a} 2D fan-beam CT geometry \textcolor{black}{was} performed. The distance from the source to the center of rotation \textcolor{black}{was} 785 mm and the distance from the source to the detector \textcolor{black}{was} 1200 mm. A circular scan \textcolor{black}{was} simulated and a total of 720 projections per rotation were acquired in an angular range of $360^0$. The detector pixel size \textcolor{black}{was} 0.388 mm and the detector \textcolor{black}{had} 1024 pixels. The dual-energy spectra \textcolor{black}{were} 100 kVp and 140 kVp, which were generated using the SpekCalc software~\citep{poludniowski2009} with 12 mm Al and 0.4 mm Sn + 12 mm Al filtration, respectively.
%Phantom description
The phantom \textcolor{black}{was} a water cylinder with inserts that \textcolor{black}{contains} a series of six solutions of varying iodine concentrations (range, 0-20 mg/mL). The diameters of the water cylinder and the inserts \textcolor{black}{were} 198 mm and 22.5 mm, respectively.

\begin{figure}[t]
    \centering
    %\vspace{-1em}
    \includegraphics[width=5in]{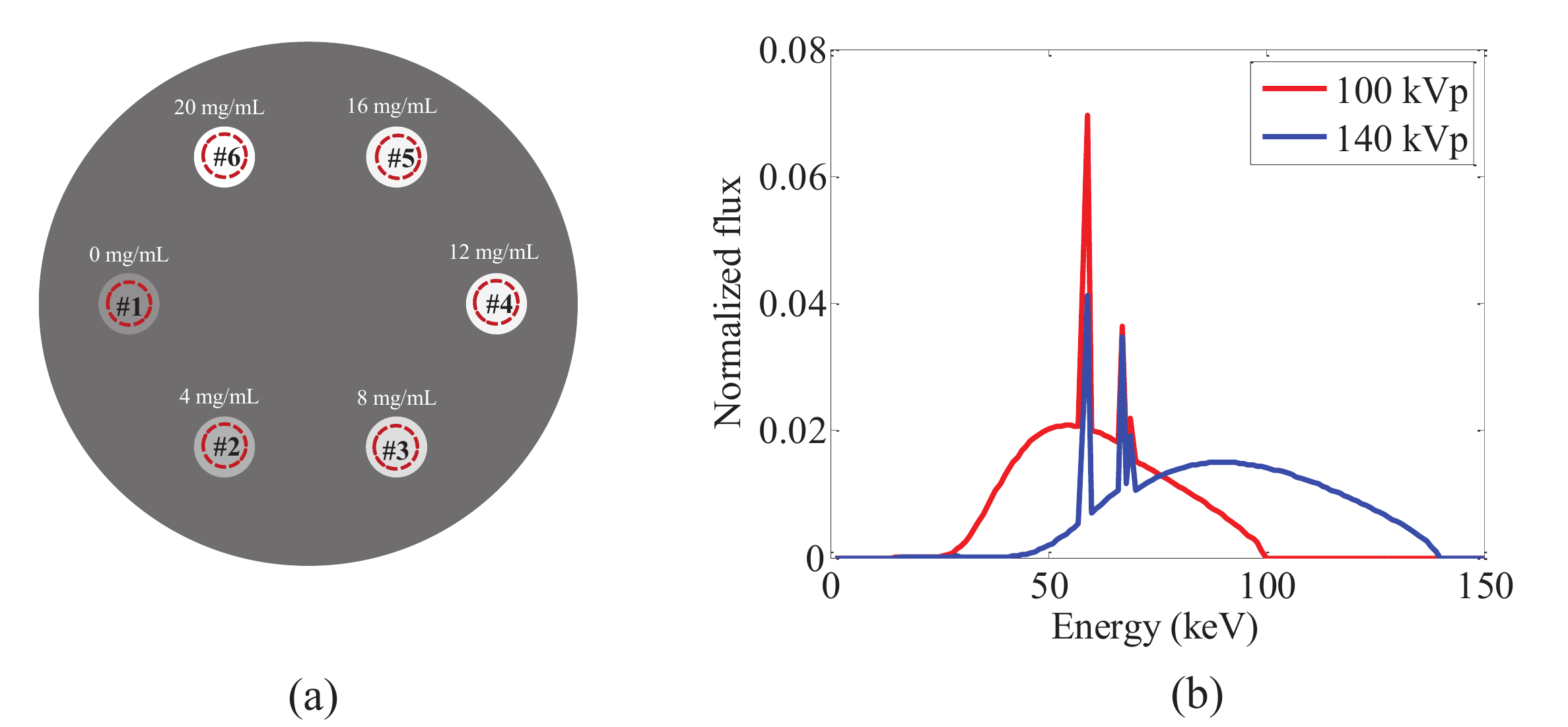}
    %\vspace{-1em}
    \caption{Iodine concentrate phantom and x-ray spectra for numerical simulation of the dual energy material decomposition. (Left) The water phantom consists of six iodine concentrate inserts. (Right) The 100 kVp and 140 kVp energy spectra used in dual energy CT. The 100 kVp spectrum was filtrated with and 12 mm Al, while the 140 kVp spectrum was filtrated with 0.4 mm Sn and 12 mm Al.}
    \vspace{-1em}
    \label{fig:f1}
\end{figure}

\begin{figure}[t]
    \centering
    %\vspace{-1em}
    \includegraphics[width=3in]{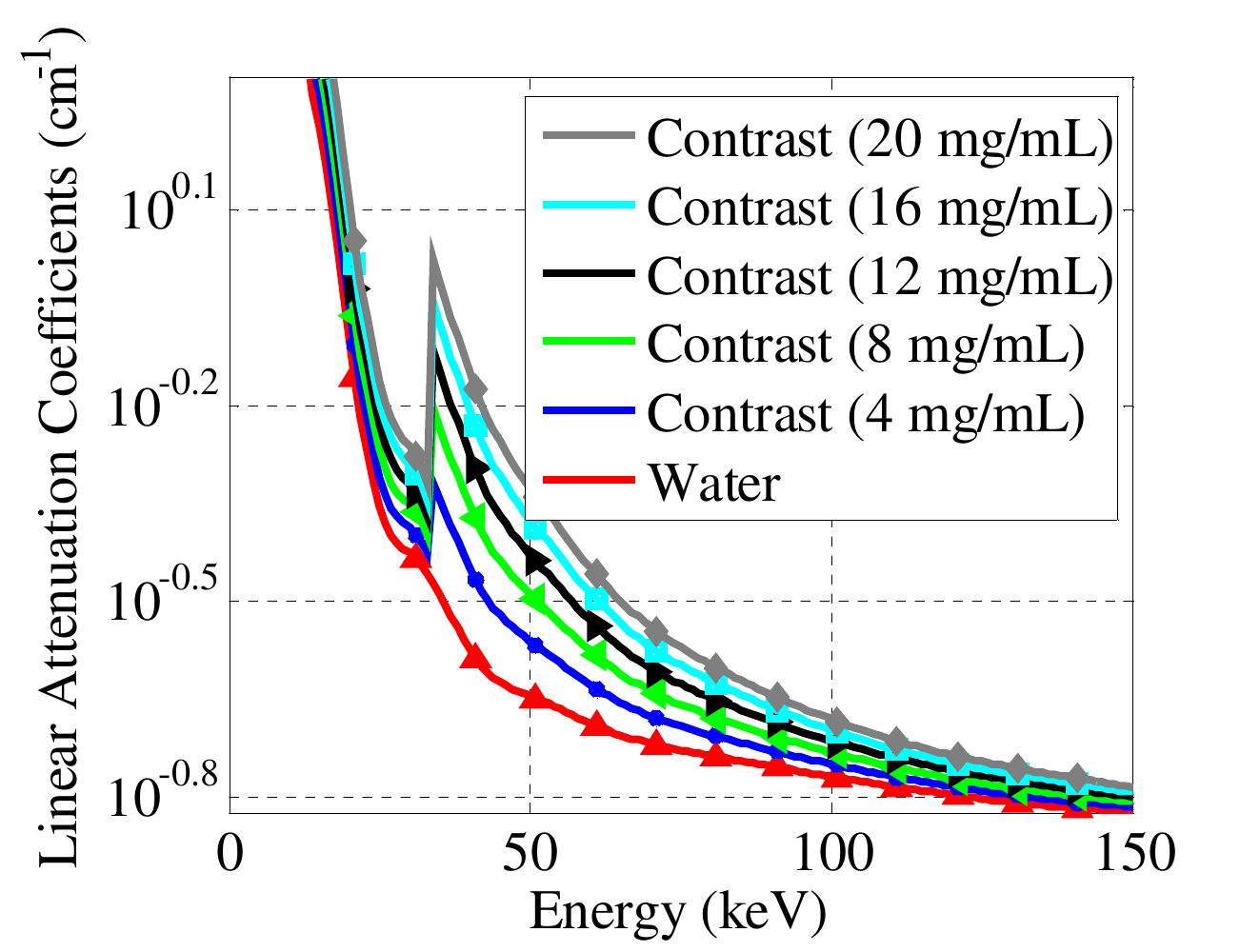}
    %\vspace{-1em}
    \caption{Linear attenuation coefficients of the iodine concentrate inserts.}
    \vspace{-1em}
    \label{fig:f2}
\end{figure}

\textcolor{black}{To be realistic, Poisson noise was considered during the numerical simulations}. \textcolor{black}{The fan beam CT projection data was created by polychromatic forward projecting the numerical phantom. Specifically, after introducing Poisson noise~\citep{nuyts2013}, the projection data can be represented as:
\beq\label{equ:weight}
I = Poisson \left\{N\int_{0}^{E_{max}}\mathrm{d}E\,\Omega(E) \, \eta(E)\,\mathrm{exp}\left[-\int_{0}^{l}\mu(E,s)\mathrm{d}s\right]\right\},
\eeq
with $N$ the total number of photons and was set to $3\times10^5$ and $1.5\times10^5$ per ray for the low- and high-energy CT scans, respectively. $\eta(E)$ \textcolor{black}{was} the energy dependent response of the detector, which was considered to be proportional to photon energy $E$ for energy-integrating detectors. $E_{max}$ \textcolor{black}{was} the maximum photon energy of the polychromatic spectrum $\Omega(E)$. $\mu(E,s)$ \textcolor{black}{was} the energy-dependent linear attenuation coefficient and \textcolor{black}{was} obtained from the National Institute of Standards and Technology (NIST) database. $l$ \textcolor{black}{was} the propagation path length for each ray and can be calculated using either analytical methods or numerical methods.} The x-ray spectra and the phantom are shown in figure~\ref{fig:f1}. The linear attenuation coefficients of the iodine concentrate inserts \textcolor{black}{calculated using the mixture rule, }are show in figure~\ref{fig:f2}. Note that the $y$ axis is plotted on a logarithmic scale. \textcolor{black}{First order beam hardening correction (water correction) was performed by simply mapping the polychromatic raysum to the corresponding monochromatic value with the incorporation of the 100 kVp and 140 kVp energy spectra.}

The calculated iodine concentrations in the phantom using different algorithms were compared with known true iodine concentrations. For comparison, direct matrix inversion without noise was also performed and the results were regarded as absolute truth. Difference image between the \textcolor{black}{absolute truth} and the images processed using different kinds of algorithms were also generated to emphasize resolution preservation and noise reduction of these algorithms.

\subsubsection{Parameters selection}

Since the parameters in the NLM filter would affect the filtered image, it \textcolor{black}{was} important to investigate the parameters selection on the final noise reduced material-specific images.
In this simulation, the search window size $S\times S$ \textcolor{black}{was} set to different values, specifically, from 11 pixels to 19 pixels with a 4 pixels interval. Material-specific images processed using HYPR-LR and HYPR-NLM methods \textcolor{black}{were} generated and compared.
Difference images between the noise-free material images and the noisy material images generated using HYPR-LR and HYPR-NLM methods, \textcolor{black}{were} calculated to depict the preservation of spatial resolution.

\subsubsection{Iteration formulation}

In order to further reduce the noise, the proposed method can be implemented in an iterative fashion, i.e., the HYPR-NLM processed material images \textcolor{black}{were} set as the input images of the next HYPR-NLM iteration. Images processed with different iteration number \textcolor{black}{were} generated and compared.

\subsubsection{Effect of dose level}

\textcolor{black}{To determine its robustness across different noise levels, we also \textcolor{black}{evaluated} the proposed HYPR-NLM method using various numbers of photons (dose levels). Numerical phantom studies using four pairs of numbers of photons ($H/L=0.5\times10^{5}/1\times10^{5}, 1\times10^{5}/2\times10^{5}, 1.5\times10^{5}/3\times10^{5}$, and $2\times10^{5}/4\times10^{5}$) \textcolor{black}{were} performed and HYPR-NLM method was employed to generate material images. For all of the simulations, the parameters of the HYPR-NLM \textcolor{black}{were} the same. Quantitative measurements including iodine concentration and noise level \textcolor{black}{were} evaluated to show the method's performance on various protocols.}

\subsection{Clinical patient data}

To evaluate the proposed method, two retrospective clinical patient studies were also performed with a dual-source DECT scanner (Somatom Definition, Siemens Healthcare).
The system acquired high- and low-energy data with two x-ray tubes with corresponding detector rings mounted onto a rotating gantry with angular offset of $90^\circ$. The two tubes (tube 1 and 2) operated independently with regard to tube voltage, tube current as well as tube filtration. Dual-energy data were acquired using the following parameters: tube 1, 140 kVp, 287 mA and tube 2, 100 kVp, 412 mA for patient 1; tube 1, 140 kVp, 345 mA and tube 2, 100 kVp, 500 mA for patient 2. The 140 kV spectrum \textcolor{black}{was} filtered with a tin filter. Images were reconstructed using the commercial software with the convolution kernel B25f for patient 1 and B26f for patient 2. All of the image slices were reconstructed in 0.75 mm slice thickness. \textcolor{black}{The matrix size for each slice \textcolor{black}{was} $512\times512$, and the pixel size \textcolor{black}{was} $0.38\times0.38$ mm$^2$.} Blending images were also generated with a linear combination (with weighting factors of 0.3 and 0.7) of the 100 kV and 140 kV images. %and 193 mm reconstruction diameter.
To obtain contrast-enhanced CT images, the contrast agent of 300 mg I/mL was injected in an antecubital vein (300 mg I/mL,Ultravist 300, Bayer HealthCare).

\textcolor{black}{The noise in the raw CT images \textcolor{black}{was} quantified using region-of-interest (ROI) analysis. Two ROIs \textcolor{black}{were} selected to calculate noise-to-signal ratio (NSR) of myocardium and ventricle, respectively.} Myocardial perfusion imaging was conducted and iodine concentration map was calculated using direct matrix inversion, HYPR-LR, Iter-DECT and HYPR-NLM. We conducted all of the evaluations on a personal laptop \textcolor{black}{which featured four Intel Core i7-4700HQ CPUs, and all of the material decomposition algorithms were implemented using MATLAB (The MathWorks, Inc., Natick, MA, United States)}. %It features four Intel Core i7-4700HQ CPUs.% and one Nvidia GeForce GTX 765M GPU.

\section{Results}
\subsection{Numerical simulations}
\begin{figure}%[t]
    \centering
    %\vspace{-3em}
    \includegraphics[width=3.5in]{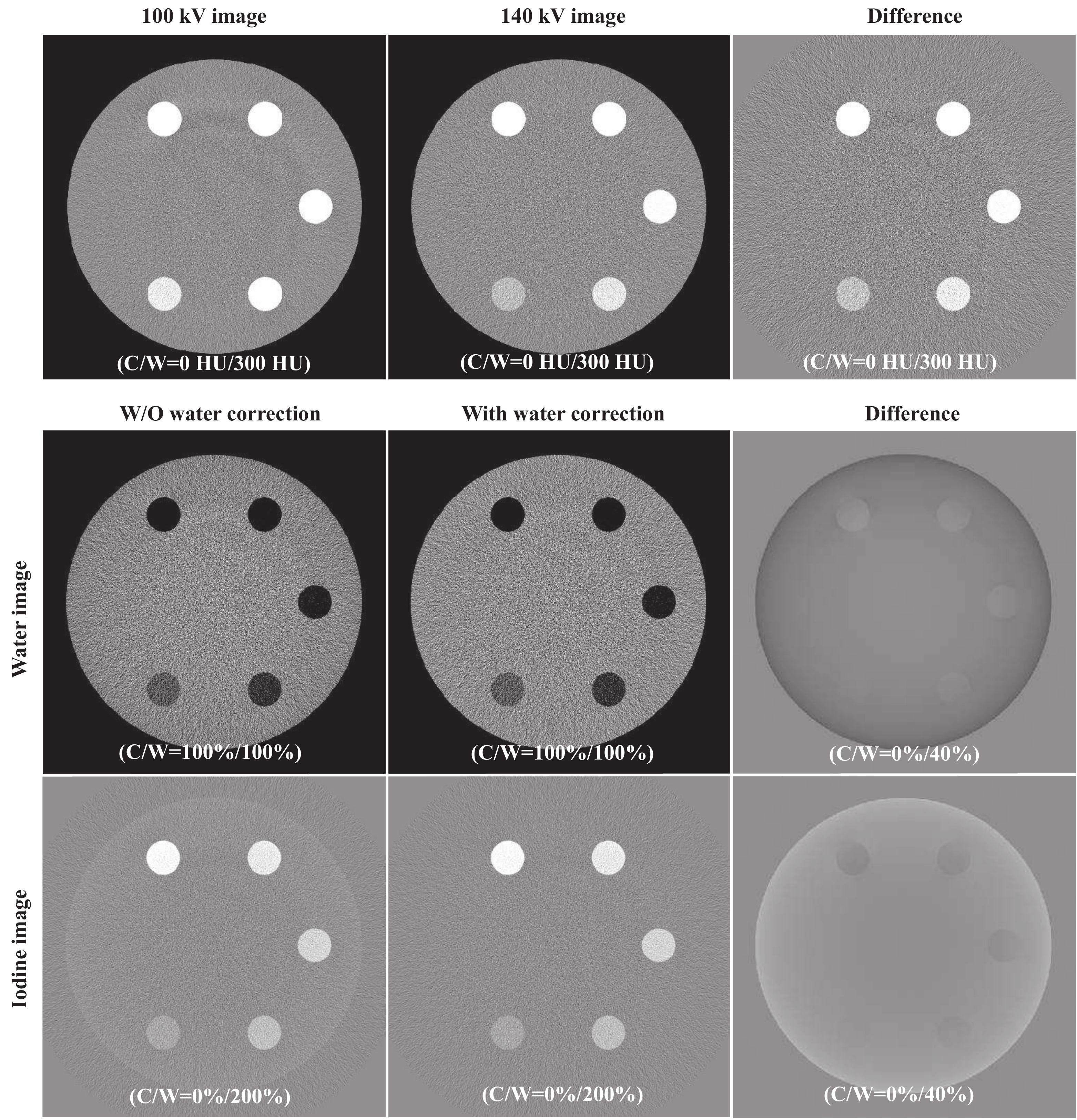}
    \caption{Reconstruction and direct dual-energy decomposition results of the numerical iodine concentrate phantom simulation. The first row shows the 100 kV and 140 kV CT images. Note that the first order beam hardening correction (water correction) has been performed for the CT images of both energies.
    Dual-energy decomposition without water correction yield cupping artifacts and inhomogeneity for the material images.}%Dispaly window: C/W=0 HU/300 HU.
    \label{fig:100kV140kVdiff}
\end{figure}

Image reconstruction results of the numerical iodine concentrate phantom are shown in figure~\ref{fig:100kV140kVdiff}. Water and the 20 mg/mL iodine concentrate are chosen as the basis materials for dual-energy material decomposition. Other iodine concentrates are expected to show up in the decomposed image corresponding to the contribution of their mass attenuation coefficients to the mass attenuation coefficient of the chosen basis material. In the first row, the 100 kV and 140 kV images and the difference between these two images are depicted. The 100 kV image shows superior iodine contrast, as expected. Note that the images of both energies have been corrected using the first order beam hardening correction algorithm and there are residual high order beam hardening artifacts (streaks) between high attenuation iodine concentrates, especially for the 100 kV image where the spectrum is much softer.

The second and the third rows show dual-energy material decomposition results using direct matrix inversion with and without water correction. As can be seen, dual-energy material decomposition yield residual cupping artifacts in the material images without water correction, and the water image and the iodine image are inhomogeneous. The difference images further clearly show the cupping artifacts and inhomogeneity. This is because the image-domain matrix inversion just linearly combine the high- and low-energy images, and \textcolor{black}{does} not take the nonlinear attenuation process of X-ray projection into account.

\begin{figure}[t]
    \centering
    %\vspace{-0.8em}
    \includegraphics[width=5in]{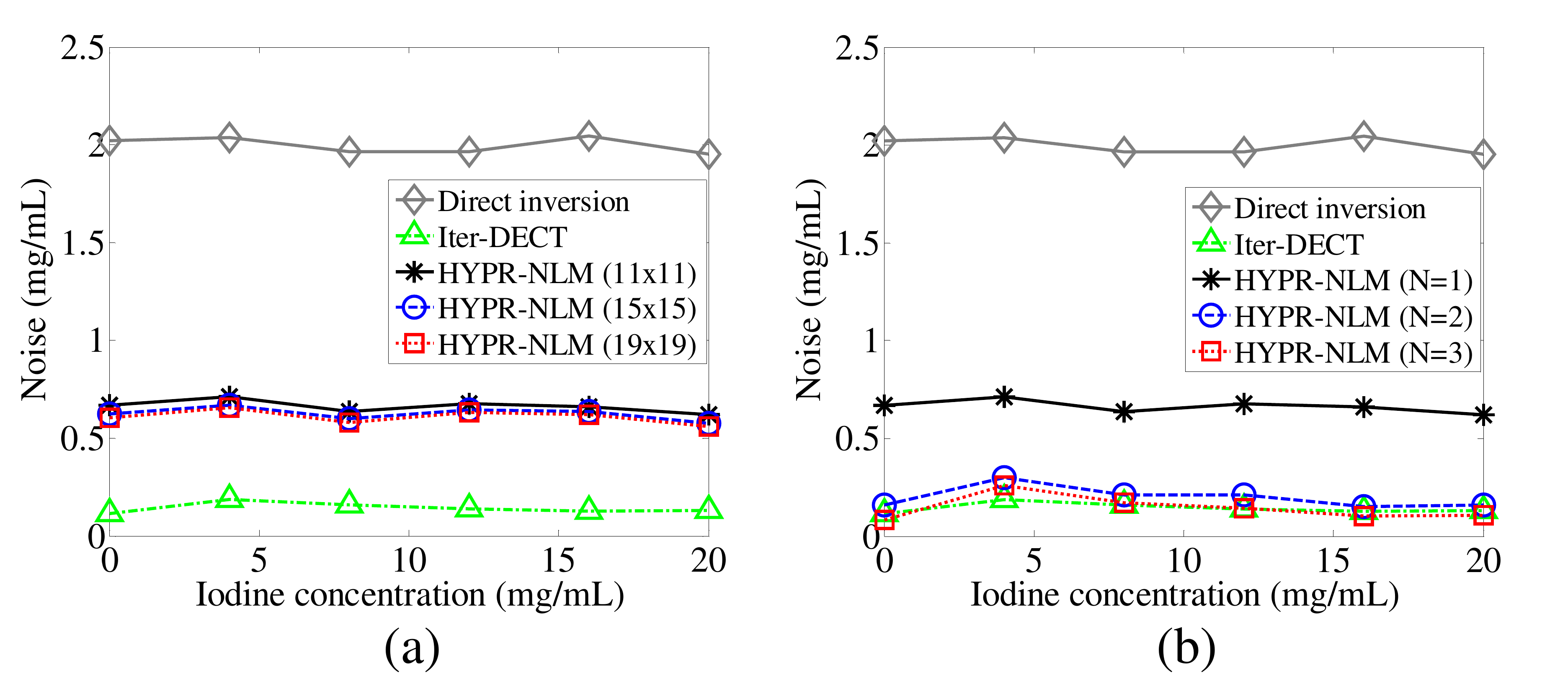}
    %\vspace{-1em}
    \caption{Iodine image noise measured at the six regions of interest for the cylinder phantom using HYPR-NLM method with different search window sizes (a) and different iteration numbers (b). For comparison, noise measured using direct matrix inversion and Iter-DECT are also depicted. }
    \label{fig:paraselection}
    \vspace{-1.5em}
\end{figure}

\begin{figure}[t]
    \centering
    \includegraphics[width=4.0in]{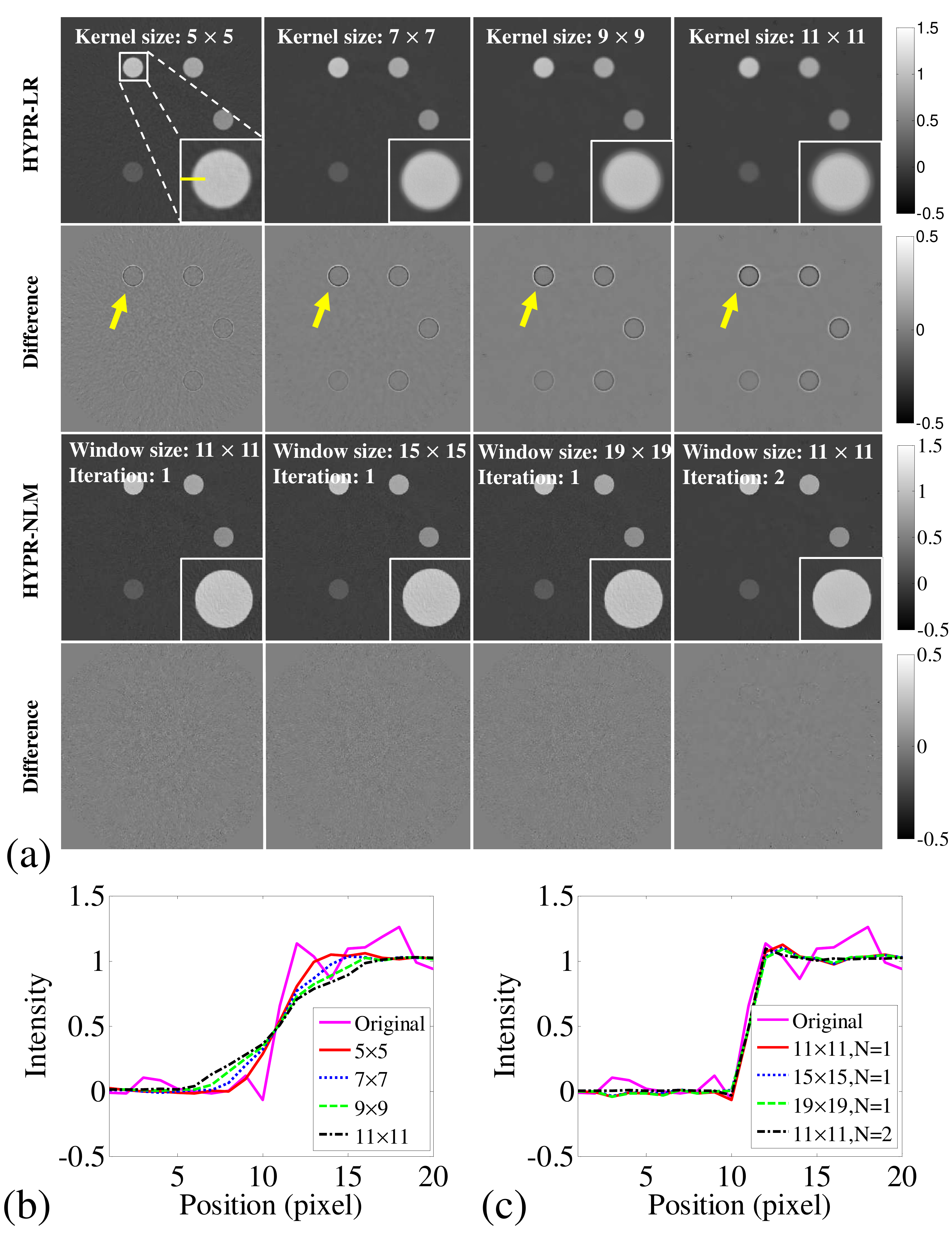}
    %\vspace{-1em}
    \caption{\textcolor{black}{Results of the numerical phantom study with respect to different parameters. Iodine images processed using HYPR-LR and HYPR-NLM with different window sizes and iteration number N (a). The difference images are the subtraction of the iodine image generated using direct inversion without noise and the individual processed iodine images. Line profiles (corresponding to the yellow line in the upper-left corner) of the iodine images processed using HYPR-LR (b) and HYPR-NLM (c) with different parameters. For comparison, line profile obtained using direct matrix inversion is also depicted (Original). }}
    \label{fig:f5}
\end{figure}

\begin{figure}[t]
    \centering
    %\vspace{-3em}
    \includegraphics[width=4.0in]{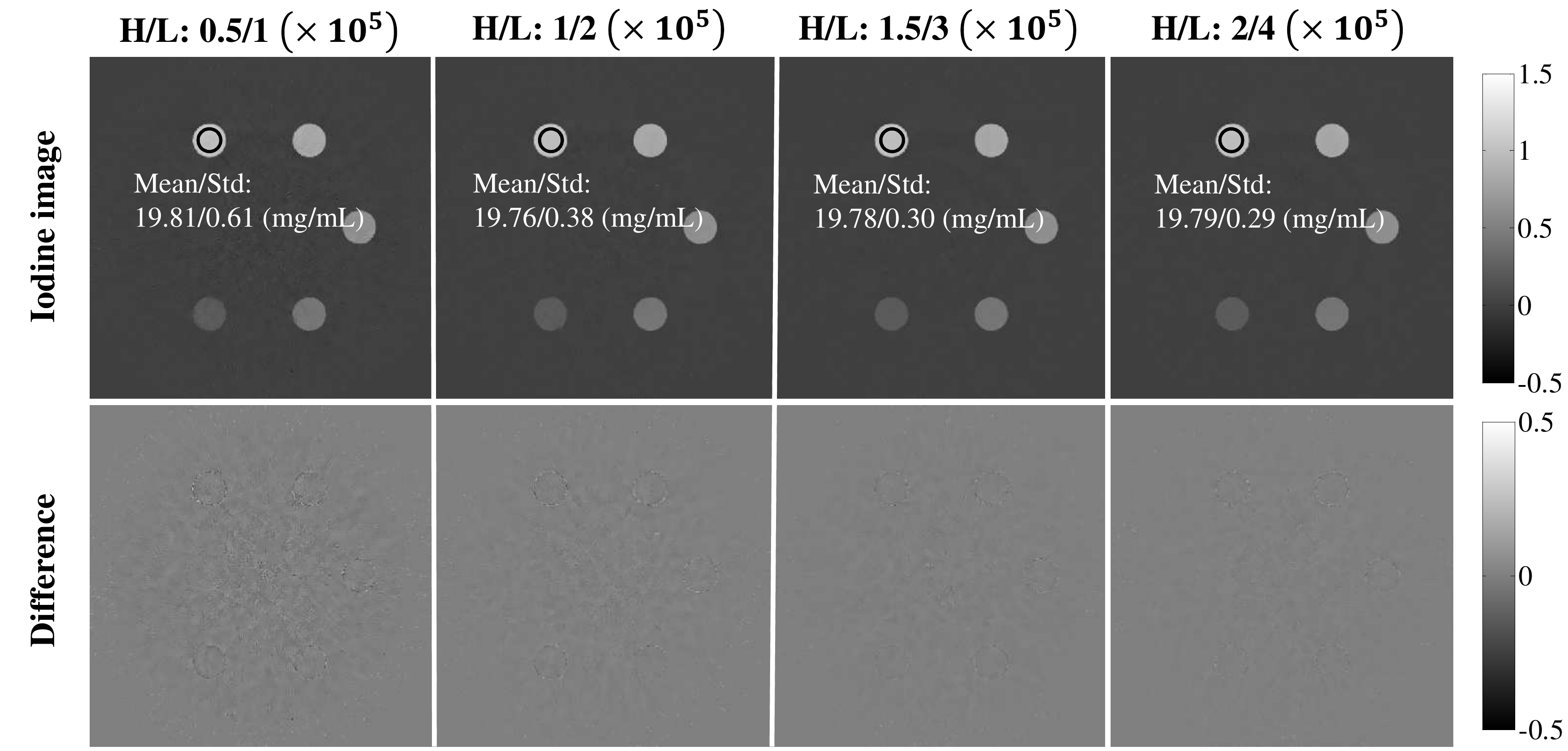}
    \caption{\textcolor{black}{Dual energy decomposed images of the iodine concentrate numerical phantom using HYPR-NLM at different numbers of photons. The difference is the subtraction of the images generated using direct inversion without noise and the HYPR-NLM images. H and L stand for the number of high- and low-energy photons per ray, respectively. The standard deviation (Std) of the ROI reduces as the number of photons increases.}}
    \label{fig:dose}
    %\vspace{-3em}
\end{figure}

Figure~\ref{fig:paraselection} shows the noise of the ROIs (labelled in figure~\ref{fig:f1}(a)) of the iodine images processed using HYPR-NLM method with different parameters. The noise reduction of HYPR-NLM with different search window sizes are depicted in figure~\ref{fig:paraselection}(a). For comparison, the results of direct matrix inversion and Iter-DECT are also present. For non-iterative HYPR-NLM formulation (i.e. the iteration number N = 1), noise magnitude is significantly reduced using HYPR-NLM method and there is no significant change between HYPR-NLM with different search windows sizes, namely, with search window size from $11\times 11$ to $19\times 19$. Thus in the following studies, we use search window size of $11\times 11$ for computational consideration. The results of noise reduction using iterative HYPR-NLM with different iterations \textcolor{black}{are} shown in figure~\ref{fig:paraselection}(b). As can been seen, with two iterations (N = 2), HYPR-NLM provides results that have comparable noise magnitude as the results of Iter-DECT method.

Figure~\ref{fig:f5}(a) shows the results of the iodine images processed using HYPR-LR and HYPR-NLM with different window sizes. Different from HYPR-LR method, which results in spatial blurring as the uniform kernel size increases from $5\times 5$ to $11\times 11$, the edges do not have significant change for the HYPR-NLM with search window sizes from $11\times 11$ to $19\times 19$ and iteration numbers from one to two. As indicated in the difference images, there are clear edge effects for the HYPR-LR method, while there are marginal edge effects for the HYPR-NLM method, suggesting the edge of the iodine images are well preserved using the HYPR-NLM method with different search window sizes. \textcolor{black}{Line profiles of the iodine images processed using HYPR-LR with different window sizes, and HYPR-NLM with different window sizes as well as iteration numbers are illustrated in figure~\ref{fig:f5}(b) and (c). These profiles clearly demonstrate HYPR-NLM can provide superior images with respect to edge-preservation, and the method is robust against parameters selection. }

\textcolor{black}{Figure~\ref{fig:dose} shows the results of dual-energy material decomposition images using HYPR-NLM method at different number of photons and quantitative analysis of the ROIs. Again, the difference image is the subtraction of the image obtained using noise-free direct matrix inversion and the HYPR-NLM image. As can be seen, HYPR-NLM works well for different scenarios. ROIs analyses indicate HYPR-NLM significantly reduces noise while preserving quantitative iodine concentration. The standard deviation of the ROI reduces as the number of photons increases. Note that H and L stand for the number of high- and low-energy photons, respectively.} % H/L stands for the number of high- and low-energy photons for each ray.

\begin{figure}[t]
    \centering
    %\vspace{-3em}
    \includegraphics[width=4.0in]{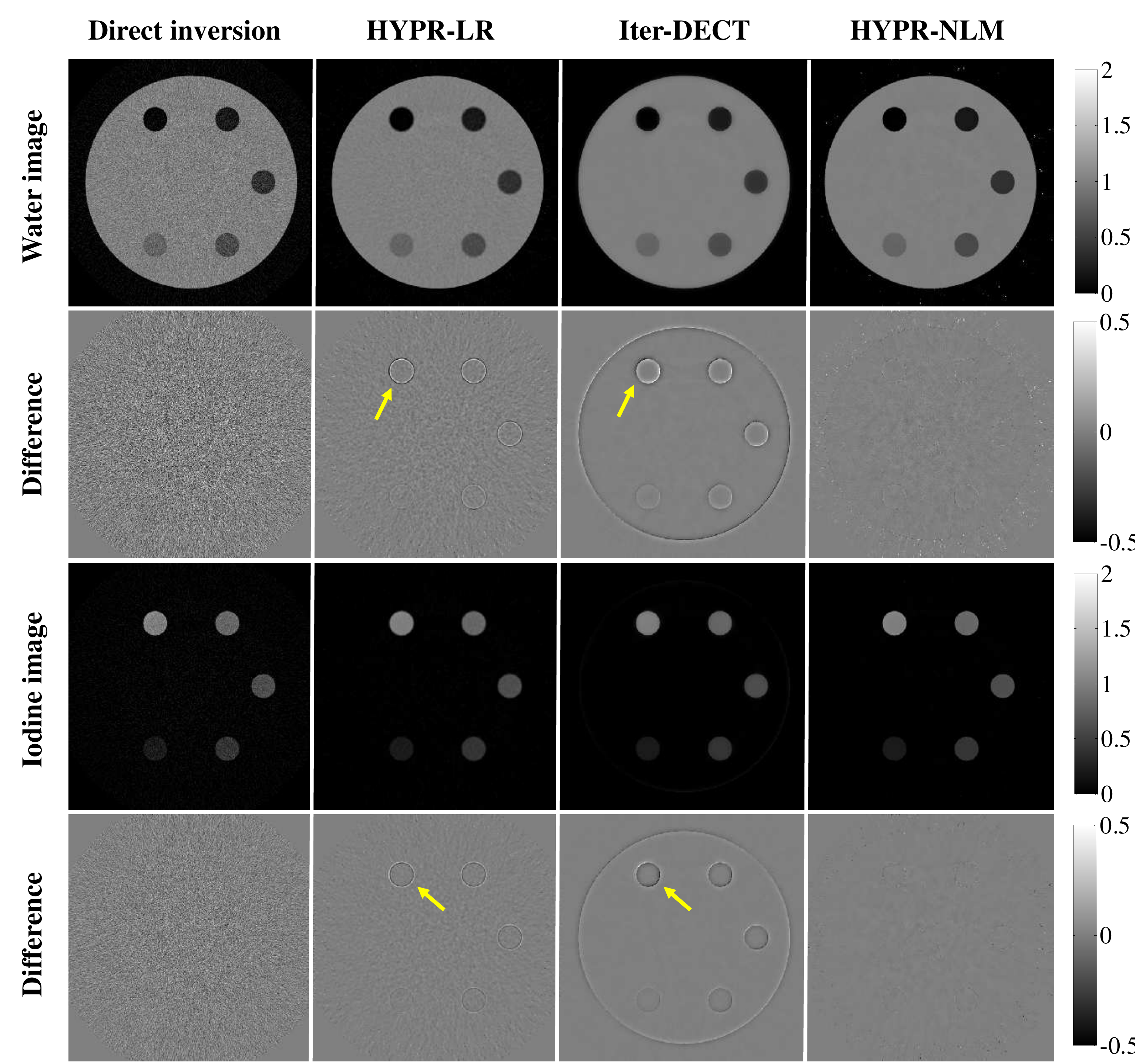}
    \caption{Dual-energy decomposed images of the iodine concentrate numerical phantom using direct matrix inversion, HYPR-LR, iterative material decomposition and HYPR-NLM. The difference is the subtraction of the images generated using direct inversion without noise and the four different methods.}
    \label{fig:f6}
    %\vspace{-3em}
\end{figure}

Material-specific images of the numerical iodine concentrate phantom obtained using direct matrix inversion, HYPR-LR, Iter-DECT and HYPR-NLM algorithms are shown in figure~\ref{fig:f6}. As can be seen, both HYPR-LR and Iter-DECT yield residual edge effects in the difference images, suggesting there are certain degree of spatial resolution degradation. However, there are marginal edge effects for the results of HYPR-NLM algorithm, showing there is minimal loss of spatial resolution. Iodine concentrations of the ROIs (labeled in figure~\ref{fig:f1}(a)) measured using different algorithms \textcolor{black}{are} shown in Table~\ref{tab:iodineConcentration}. Compared to the true values of the iodine concentrate inserts, the quantitative \textcolor{black}{measurements} of iodine concentration using the four algorithms \textcolor{black}{are} well preserved. %HYPR-NLM outperforms HYPR-LR while Iter-DECT yields the lowest noise level.

\begin{table}
\caption{Iodine concentration measured using different image-domain dual-energy material decomposition algorithms and true iodine concentration in phantom.}
\label{tab:iodineConcentration}
\begin{center}
\begin{tabular}{p{0.15\textwidth}<{\centering}p{0.15\textwidth}<{\centering}
p{0.15\textwidth}<{\centering}p{0.15\textwidth}<{\centering}p{0.15\textwidth}<{\centering}} %% this
\toprule
\rule[-1ex]{0pt}{3.5ex}   True value (mg/mL) & Direct inversion (mg/mL)& HYPR (mg/mL)& Iter-DECT (mg/mL)& HYPR-NLM (mg/mL)\\
\hline
\rule[-1ex]{0pt}{3.5ex}   0 & 0.10 & 0.10 & 0.12 & 0.10  \\
\rule[-1ex]{0pt}{3.5ex}   4 & 4.24 & 4.26 & 4.24 & 4.26  \\
\rule[-1ex]{0pt}{3.5ex}   8 & 8.36 & 8.36 & 8.34 & 8.36  \\
\rule[-1ex]{0pt}{3.5ex}   12 & 12.26 & 12.24 & 12.22 & 12.24  \\
\rule[-1ex]{0pt}{3.5ex}   16 & 16.08 & 16.10 & 16.06 & 16.10  \\
\rule[-1ex]{0pt}{3.5ex}   20 & 19.94 & 19.94 & 19.92 & 19.94  \\
\bottomrule
\end{tabular}
%\bottomrule
\end{center}
%\vspace{-2em}
\end{table}

%\subsubsection{Parameters selection}
%\subsubsection{Iteration formulation}
%\subsection{Comparison to non-local mean filter}
%\vspace{-3mm}
\subsection{Patient study}

\begin{figure*}[t]
    \centering
    \includegraphics[width=0.6\textwidth]{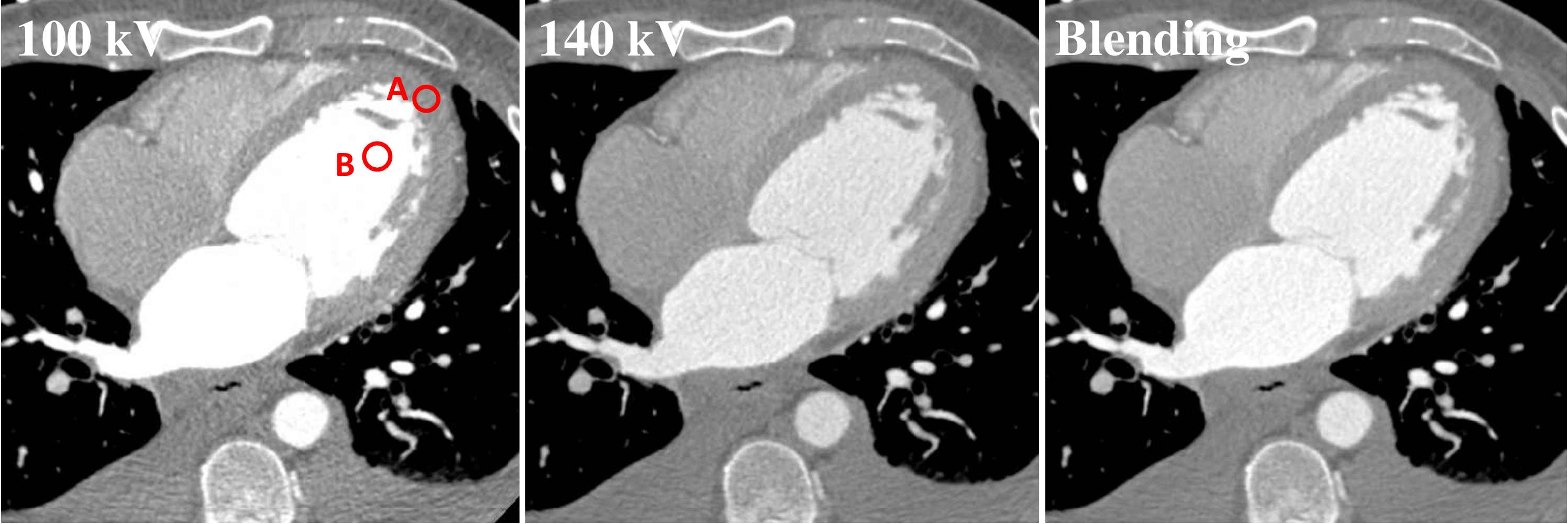}
    \caption{Patient 1, 100 kV (left), 140 kV (middle) and blending (right) CT images of myocardial perfusion imaging using dual-source DECT scanner (C = 0 HU, W = 1000 HU). \textcolor{black}{The two region-of-interests A and B (red circles) are used to calculate signal-to-noise ratio of myocardium and ventricle, respectively.}}% the difference between the high- and low-energy images demonstrate motion artifacts.
    \label{fig:f7}
\end{figure*}

\begin{figure*}[t]
    \centering
    \includegraphics[width=0.72\textwidth]{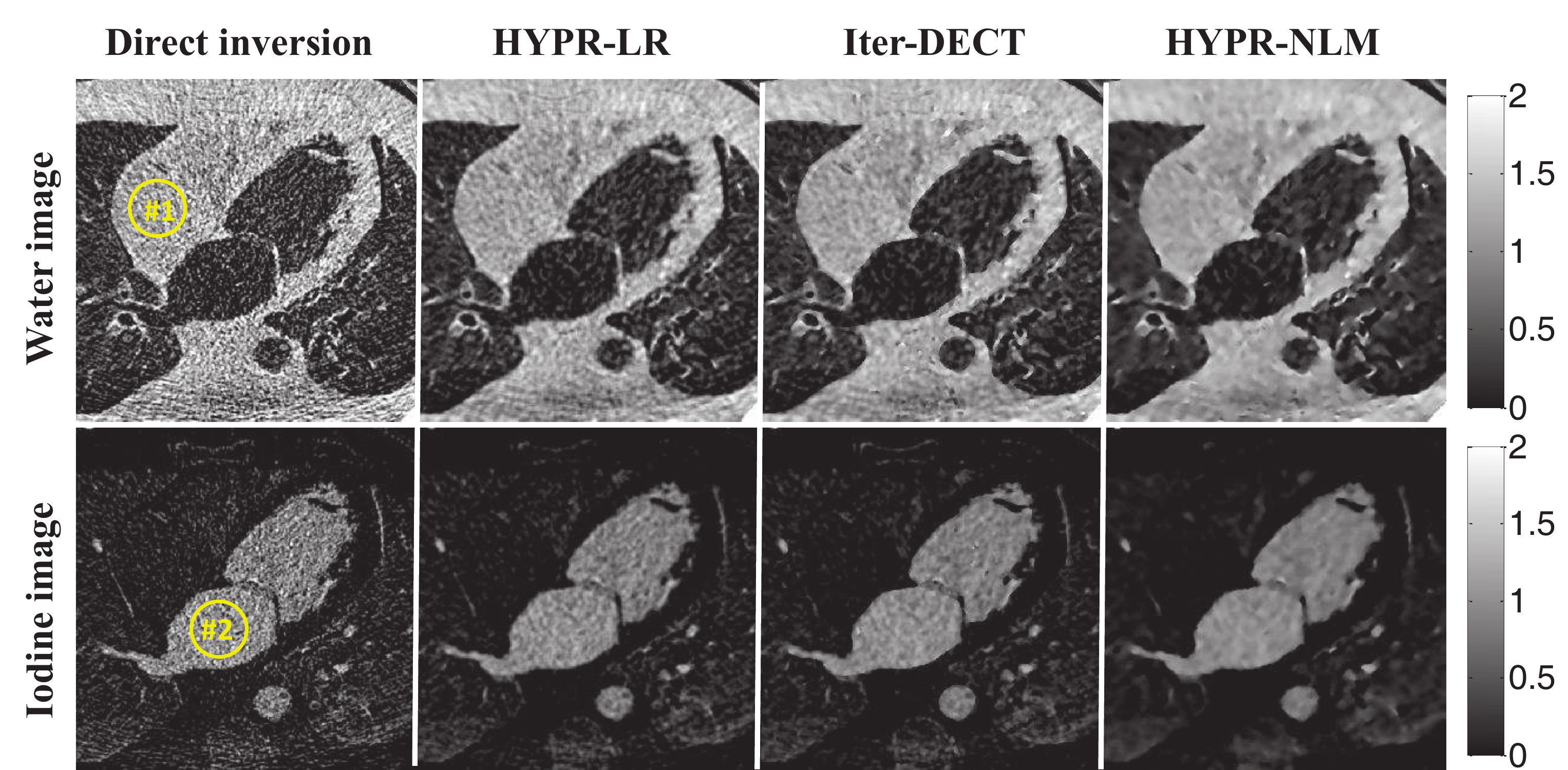}
    \caption{Patient 1, dual-energy decomposed images of myocardial perfusion imaging using direct matrix inversion, HYPR-LR, Iter-DECT and HYPR-NLM.}
    \label{fig:f8}
\end{figure*}

Figure~\ref{fig:f7} shows the low-energy (100 kV), high-energy (140 kV) CT images and \textcolor{black}{their} linear blending images \textcolor{black}{with weighting factors of 0.3 and 0.7 \textcolor{black}{for} the myocardial perfusion imaging of patient 1. Two regions-of-interests (ROI) A and B are labelled respectively on myocardium and ventricle to calculate their NSRs. The NSRs of the myocardium for the 100 kV and 140 kV CT images are $34.6\%$ and $27.9\%$, respectively. While the NSRs of the ventricle for the 100 kV and 140 kV CT images are $5.3\%$ and $5\%$, respectively. For both ROIs, the NSRs of the 100 kV images are higher than the 140 kV images. This may be the reason why the 140 kV image has a higher weighting factor in the blending image. }Water image, and iodine contrast image decomposed using the high- and low-energy images with different algorithms are present in figure~\ref{fig:f8}. As can be seen, dual-energy material decomposition using direct matrix inversion yields \textcolor{black}{increased} noise. HYPR-LR reduces the noise level to a certain extent. Iter-DECT significantly reduces the amplified noise. For both of the water and iodine images, HYPR-NLM \textcolor{black}{shows} superior image quality. Quantitative \textcolor{black}{measurements} of iodine concentration and noise reduction of the labelled ROIs are depicted in table~\ref{tab:patientData}. %however, it also shows patchy effect. Figure~\ref{fig:f7} (a), (b) and (c) show the low-energy 100 kV, 140 kV and, respectively.

\begin{figure*}%[t]
    \centering
    \includegraphics[width=0.6\textwidth]{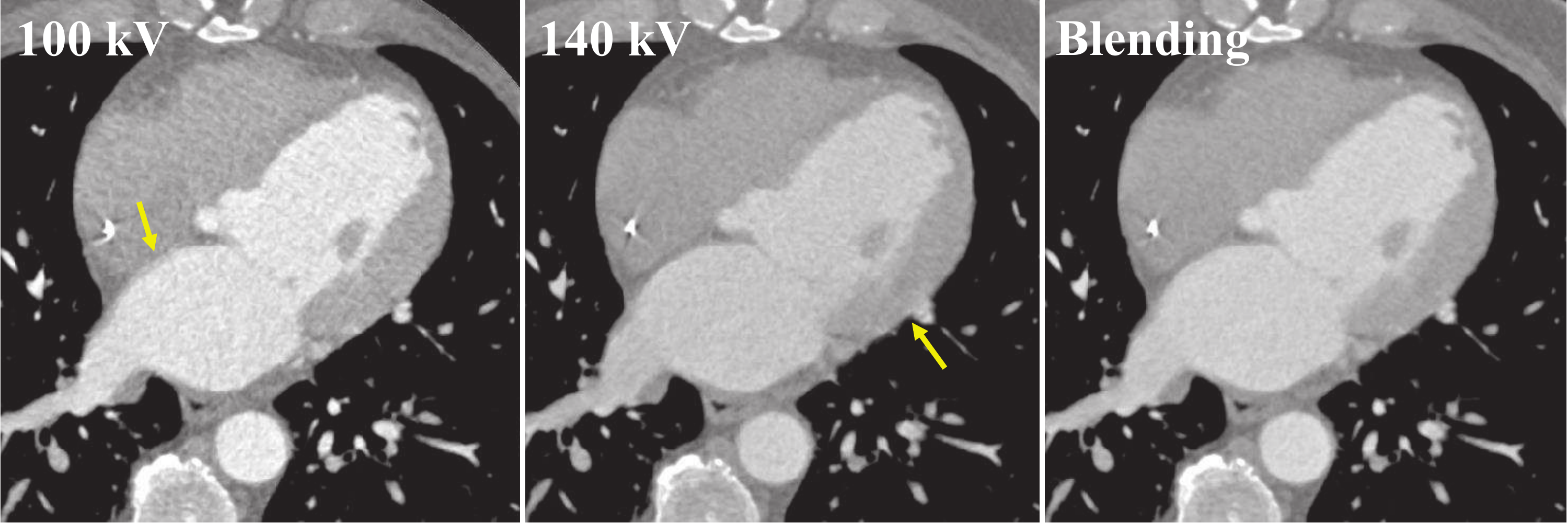}
    \caption{Patient 2, 100 kV (left), 140 kV (middle) and blending (right) CT images of myocardial perfusion imaging using dual-source DECT scanner (C = 0 HU, W = 1000 HU). }
    \label{fig:f9}
\end{figure*}

\begin{figure*}[bt]
    \centering
    %\vspace{-20mm}
    \includegraphics[width=0.72\textwidth]{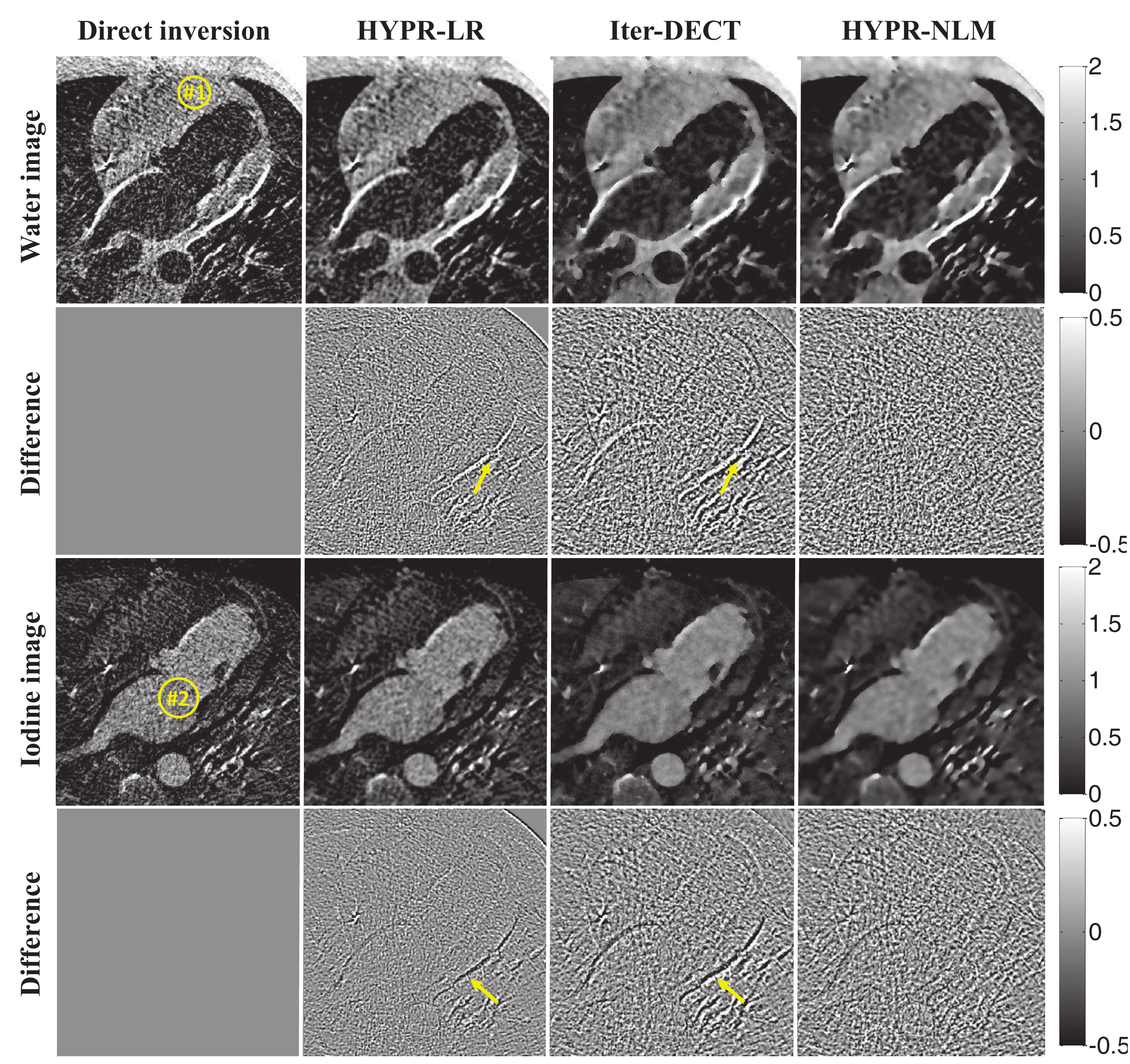}
    \caption{Patient 2, dual-energy decomposed images of myocardial perfusion imaging using direct matrix inversion, HYPR-LR, Iter-DECT and HYPR-NLM. The difference is the subtraction of the direct inversion and the other methods.}
    \label{fig:f10}
\end{figure*}

The images of patient 2, a case with motion artifact presented in figures~\ref{fig:f9}-\ref{fig:f10}, provides a challenging situation for dual-energy material decomposition. \textcolor{black}{This is because myocardial imaging is often compromised by motion blur, leading to DECT artefacts.} The blurring edge labelled in the 100 kV image of figure~\ref{fig:f9} does not \textcolor{black}{have a} consistent profile \textcolor{black}{compared to} the 140 kV image, which shows a sharper edge \textcolor{black}{here}. This happens for the labelled edge for the 140 kV image as well. Again, the blending image shows the linear combination of the 100 kV and 140 kV CT images.

\textcolor{black}{Water} images and iodine contrast images, as shown in figure~\ref{fig:f10}, clearly depict the error induced by motion. Still, there are marginal edge effects for HYPR-NLM processed images, suggesting improvements compared to HYPR-LR and Iter-DECT. Dual-energy decomposition using HYPR-NLM leads to more homogeneous results and more significant noise reduction.

Table~\ref{tab:patientData} shows the mean values and standard deviations of the ROIs measured using direct matrix inversion, HYPR-LR, Iter-DECT and HYPR-NLM algorithms. \textcolor{black}{Note that the NSR for direct matrix inversion is approximately $30\%$ for both basis material images and is reduced by factors for approximately 2, 3 and 5 for HYPR-LR, Iter-DECT and HYPR-NLM, respectively.} The computational time for \textcolor{black}{each slice} per iteration for HYPR-NLM is about 280 s, compared to 300 s for Iter-DECT. There are no significant differences between the mean values measured using the four algorithms, suggesting the quantitative measurements of iodine concentration are acceptable for all four approaches. For noise reduction, direct inversion and HYPR-NLM yield the highest and the lowest noise level, respectively. HYPR-NLM outperforms HYPR-LR in both resolution preservation and noise reduction.

\begin{table*}
%\vspace{-5mm}
\caption{The mean values and standard deviations of the ROIs (labeled as \#1 and \#2 in figure~\ref{fig:f8},~\ref{fig:f10}) of the myocardial data.}
\vspace{-1.5mm}
\label{tab:patientData}
\begin{center}
%\begin{tabular}{cp{0.07\textwidth}<{\centering}p{0.07\textwidth}<{\centering}
%p{0.07\textwidth}<{\centering}p{0.07\textwidth}<{\centering}p{0.07\textwidth}<{\centering}} %% this
\begin{tabular}{cccccc}
\toprule
%\multicolumn{4}{c}{Patient 1} & \multicolumn{4}{c}{Patient 2} \\ \cmidrule(r){2-5} \cmidrule(l){6-9}
 &    & Direct inversion & HYPR-LR & Iter-DECT & HYPR-NLM \\
\hline
\multicolumn{1}{ c }{\multirow{2}{*}{Patient 1} } & Water image & 1.16$\pm$0.36 & 1.16$\pm$0.16 & 1.16$\pm$0.12 & 1.16$\pm$0.08 \\& Iodine image & 0.99$\pm$0.31 & 0.99$\pm$0.14 & 0.99$\pm$0.12 & 0.99$\pm$0.08 \\
\hline
\multicolumn{1}{ c }{\multirow{2}{*}{Patient 2} } & Water image  & 1.04$\pm$0.34 & 1.04$\pm$0.18 & 1.04$\pm$0.08 & 1.04$\pm$0.07\\
&Iodine image & 0.97$\pm$0.28 & 0.98$\pm$0.14 & 0.98$\pm$0.06 & 0.98$\pm$0.05 \\
%\rule[-1ex]{0pt}{3.5ex}  Water image & 1.16$\pm$0.36 & 1.16$\pm$0.16 & 1.16$\pm$0.12 & 1.16$\pm$0.08& 1.04$\pm$0.34 & 1.04$\pm$0.18 & 1.04$\pm$0.08 & 1.04$\pm$0.07\\
%\rule[-1ex]{0pt}{3.5ex}  Iodine image & 0.99$\pm$0.31 & 0.99$\pm$0.14 & 0.99$\pm$0.12 & 0.99$\pm$0.08 & 0.97$\pm$0.28 & 0.98$\pm$0.14 & 0.98$\pm$0.06 & 0.98$\pm$0.05 \\
\bottomrule
\end{tabular}
\end{center}
\end{table*}

%\vspace{-0.3em}
%%%%%%%%%%%%%%%%%%%%%%%%%%%%%%%%%%%%%%%%%%%%%%%%%%%%%%%%%
\section{Discussion and Conclusion}%%%%%%%%%%%%%%%%%%%%%%%%%%%%%%
%%%%%%%%%%%%%%%%%%%%%%%%%%%%%%%%%%%%%%%%%%%%%%%%%%%%%%%%%%
\label{sec:discussion}

HYPR-NLM has shown to provide sensible results for noise reduction of image-domain dual-energy material decomposition. For numerical simulation studies, where there exists \textcolor{black}{certainty} (noiseless material images obtained using direct matrix inversion), HYPR-NLM outperforms both HYPR-LR and Iter-DECT for spatial resolution preservation. For the clinical myocardial perfusion imaging studies, to show the edge preservation of the dual-energy material-decomposition algorithms, material-specific images are subtracted from the noisy images generated using direct inversion to see whether there are any noticeable anatomical features present.

\textcolor{black}{We have demonstrated that the HYPR framework can be applied directly to the basis material images. The initial basis material images can be generated using matrix inversion because this method performs material decomposition without compromising spatial details, which can be exploited by the HYPR-NLM algorithm to yield both noise reduced and spatial information well preserved material images. To achieve this goal, it is also possible to filter the DECT images before the application of material decomposition, or directly apply a feature preserved noise reduction filter on the material decomposition images~\citep{cai2015}. In the future, comprehensive comparative studies of these methods will be performed after sufficient experiences and data are accumulated. }%we would like to perform comparison studies via systematic investigation of these methods.

For the HYPR-LR algorithm, an uniform kernel is employed to convolve with the energy image and the composite image, thus anatomical structures of the local neighborhood are introduced into the processed image. When the window size of the uniform kernel increases, more local features are introduced into the convolution, resulting spatial information degraded material-specific images, as depicted in the first row of figure~\ref{fig:f5}. To the contrary, for the HYPR-NLM algorithm, the non-local mean is employed in the convolution procedure. The resulting image is a weighted average of all pixels in the original image, where the weight is determined by similarity between two pixels and the similarity is a measurement of the geometrical configuration in a square neighborhood $\Theta$. Thus pixels with a similar image value in \textcolor{black}{their neighborhoods} have larger weights and contribute more in the whole image averaging, while local pixels may have smaller weights and consequently contribute less if their geometrical configurations are not similar to the targeting pixel. In this case, since all of the image pixels can contribute to the targeting pixel according to their similarities, few local features are introduced and spatial information is well preserved in the final material image. In addition, the similarity depended weights make HYPR-NLM robust with respect to the size of averaging window (the search window), as indicated in the third row of figure~\ref{fig:f5}. It has to note instead of using the whole image, we have restricted the search windows in size of $11\times11, 15\times15, 19\times19$ pixels for computational purpose.

%For DECT imaging, the low- and high-energy CT images have geometrical self-similarities. The composite image has lower noise and also has geometrical self-similarities with the energy images.depends on the noise level of the composite image which %Noise reduction using HYPR-LR framework can exploit the information redundancies in the energy domain.
\textcolor{black}{For DECT imaging, the low- and high-energy CT images have geometrical self-similarities, which have been widely used in dynamic tomography reconstruction. When the HYPR-LR algorithm is applied to the CT images, the self-similarities are exploited by the composite image which has lower image noise level than the individual CT images. In this sense, the energy dimension is regarded as the time dimension in four-dimensional CT. The noise of HYPR-LR processed image mainly depends on the noise level of the composite image. For the HYPR-NLM method, the noise of the processed image does not follow the rule from two aspects: (1) Different from original HYPR-LR method that is applied to CT images, HYPR-NLM is directly applied to the material images; (2) the weights $\omega(i,j)$ used to filter the basis material image and composite image are different, while HYPR-LR uses the same kernel for both CT image and composite image. In addition, when it is performed in iterative formulation, the noise level of material image obtained using HYPR-NLM can be further reduced, as clearly demonstrated by the numerical simulation and clinical patient studies.} We have found two consecutive HYPR-NLM calculations could yield satisfactory results. Note that the algorithm is not optimized for time consideration, thus the computational time can be further reduce via algorithm optimization and parallel acceleration.%For the HYPR-NLM algorithm,  is comparable to, or even lower than that generated using Iter-DECT.

%\section{Summary}
In summary, the proposed HYPR-NLM algorithm incorporates the edge-preserving non-local mean into the HYPR-LR framework and provides an effective way to suppress the noise magnification in material decomposition which has been a generic problem in DECT. A comparison of the technique with direct matrix inversion, and with published HYPR-LR as well as image-domain material decomposition algorithms suggests that all four algorithms yield acceptable quantitative measurement of iodine concentration. Direct matrix inversion yields the highest noise level, followed by HYPR-LR and Iter-DECT. HYPR-NLM significantly reduces noise while preserving the accuracy of quantitative measurement and spatial information.

\ack
This work is supported in part by NIH grants 7R01HL111141 and 1R01-EB016777. This work is also supported by the Natural Science Foundation of China (NSFC Grants No. 81201091 and No. 81171402), the 863 plan of the Ministry of Science and Technology of China (Grant No. 2015AA020917), Fundamental Research Funds for the Central Universities in China, Fund Project for Excellent Abroad Scholar Personnel in Science and Technology and Guangdong innovation team of image-guided Therapy (No. 2011S013).
%\vspace{-3.2mm}
\appendix
\section{Iterative image domain material decomposition}
For comparison, an iterative image domain dual-energy material decomposition method which significantly \textcolor{black}{reduced} the \textcolor{black}{increased} material decomposition noise, was also introduced. This method \textcolor{black}{balanced} the data fidelity of image value of direct inversion material decomposition and quadratic error of decomposed images using an optimization framework~\citep{niu2014}. It \textcolor{black}{was} referred to as Iter-DECT in this work. The optimization problem is formulated as follows,
\beq\label{equ:costfunction}
min_{\vec{x}} F(\vec{x})=(A\vec{x}-\vec{\mu})^{T}V^{-1}(A\vec{x}-\vec{\mu}) + \lambda\cdot R(\vec{x}),
\eeq
where $R$ -- the quadratic penalty term; $\lambda$ -- the constant to adjust the relative weights between the data fidelity term and the smooth term. The penalty term is defined as follows,
\begin{eqnarray}
%\beq
R(\vec{x}) = \frac{1}{2}\underset{\mathbf{i}}\sum \underset{\mathbf{k\in N_i}} \sum e_{ik} (x_i-x_k)^2,
\end{eqnarray}
with $N_i$ the set of four nearest neighbors of the $i$th pixel in the image and $e_{ik}$ the edge-detection weight for pixel $i$ and $k$. %The above equation can be explicitly written as

Nonlinear conjugate gradient (CG) method \textcolor{black}{was} used to minimize the cost function defined by equation~(\ref{equ:costfunction}).
During CG iterations, the gradient \textcolor{black}{was} calculated by the partial derivation of the cost function with respect to $\vec{x}$,
\beq\label{equ:CG problem}
\nabla_{\vec{x}} F(\vec{x})= 2 A^{T}V^{-1}(A\vec{x}-\vec{\mu}) + \lambda\cdot \nabla R(\vec{x}),
\eeq
with $\nabla R = (\frac{\partial R}{\partial x_1}, \frac{\partial R}{\partial x_2}, \cdots \frac{\partial R}{\partial x_{2N}})$. %and

%\begin{acknowledgments}
% Put your acknowledgments here.
%\end{acknowledgments}

%novelty
%Magnified noise is a general concern for dual-energy material decomposition. In this study, we develop an image-domain material decomposition algorithm for dual-energy CT (DECT). Comparison studies demonstrate the proposed algorithm significantly reduce the DECT material decomposition noise while preserving quantitative measurements and high-frequency edge information. In addition, the method is robust with respect to parameters. It may significantly improve DECT imaging by providing quantitative iodine concentration, which plays a important role in myocardial perfusion imaging and other clinical applications.

%\bibliography{dualenergyCT}
%\begin{thebibliography}{10}
%\begin{harvard}

%\end{thebibliography}

%\end{harvard}

\end{document}